\documentclass{article}
\usepackage[utf8]{inputenc}
\usepackage{graphicx,amsmath,amsfonts}
\usepackage{amsthm,graphicx,tikz,float}
\usepackage{enumerate,array}
\usepackage{pgfplots,relsize,color}
\usepackage{times,lipsum}
\usepackage{xurl,float}
\usepackage{caption,enumitem}
\pgfplotsset{compat=1.18}

\graphicspath{ {./Figures/} }

\usetikzlibrary{arrows,decorations,backgrounds}
\usepgfplotslibrary{fillbetween}

\newtheorem{lemm}{Lemma}
\newtheorem{thm}{Theorem}
\numberwithin{thm}{section}

\numberwithin{eg}{section}

\newcounter{comments}
\definecolor{Red}{rgb}{0.8,0,0}
\definecolor{Blue}{rgb}{0,0,0.8}

\definecolor{darkpink}{RGB}{219, 48, 122}

\begin{document}
\title{Invasive species control via a discrete model for the Trojan Y-chromosome strategy}
\maketitle

\centerline{\scshape  Don Kumudu Mallawa Arachchi$^{1}$, Rana D. Parshad$^{1}$ and  Claus Kadelka$^{1}$}
\medskip

    \centerline{ 1) Department of Mathematics,}
 \centerline{Iowa State University,}
   \centerline{Ames, IA 50011, USA.}


\begin{abstract}
Invasive species are a growing threat to ecosystems, particularly in aquatic environments. The Trojan Y Chromosome (TYC) strategy is a promising biological method for reducing invasive populations by introducing genetically modified males (supermales) that produce only male offspring, leading to population decline due to a shortage of females. In this study, we develop a novel discrete-time, age-structured mathematical model to simulate the effects of this strategy. Our model divides the species' life cycle into two stages—egg and maturity—and tracks different sub-populations, including supermales. We analyze the model’s equilibria and prove the existence and stability of extinction and positive equilibrium points. Numerical simulations show that extinction depends on factors such as fecundity, the number of supermales released, and initial population sizes. The model also reveals complex behaviors, such as bistability and thresholds for population collapse. This discrete approach offers a useful framework for understanding and optimizing the TYC strategy and can help guide future field applications of invasive species control.
\end{abstract}

\section{Introduction}
An invasive species is defined as a non-native organism that has been introduced to a new area and is likely causing ecological and economic damage \cite{myers2000eradication, arim2006spread}.
The rate of alien invasions continues to increase worldwide at an unprecedented rate \cite{seebens2017no, diagne2021high}, and marine invasions form a large part of this \cite{schofield2015non, dunham2020genetically}. Marine invaders have led to large-scale hindrance to marine biodiversity, disruption of ecosystem structure and functions, created vectors for disease as well as causing loss to native species populations and habitats~\cite{lovell2006economic, charles2007impacts}. Thus, the control of aquatic invaders is a paramount issue in ecology. Chemical and biological controls constitute the main strategies used today \cite{bampfylde2007biological}. Chemical control is the use of toxic chemical substances such as pesticides (e.g., rotenone), administered at regular intervals to directly reduce alien species density. These chemicals are however environmentally harmful, and repeated applications of a pesticide can cause the invasive species to become resistant \cite{ling2003rotenoneoa}. This motivates the search for  non-toxic and self-sustaining approaches \cite{teem2020genetic,perrin2009sex}. 

A typical case of biological control is the introduction of natural enemies or predators released to target the invasive species via direct depredation. However certain ``out of the box" versions have also been introduced. These fall under the broad umbrella of genetic bio-control. One such popular method is the Trojan Y chromosome strategy (TYC), which was originally proposed by Gutierrez and Teem \cite{gutierrez2006model, gutierrez2012analysis} to control invasive species with XX - XY sex chromosomes via a constant release of YY feminized males. Mating with this feminized male will only lead to male progeny. This will skew the sex ratio over time and hence cause local extinction due to the absence of females. Various versions of this strategy exist: (i) YY males could be continuously introduced into an invasive population \cite{parshad2011global, parshad2011long}, or (ii) the strategy could be mimicked via male stocking and female harvesting \cite{lyu2020comparison}. Notably, this strategy is now in field practice in the United States \cite{diagne2021high}. It is employed in Idaho \cite{dunham2020genetically}, and also in New Mexico for the control of invasive brook trout (\emph{Salvelinus fontinalis}) \cite{dunker2022operational}. It is also in early use by several state agencies in Alaska to control invasive northern pike (\emph{Esox lucius}) \cite{dunker2022operational}. This is part of a large scale movement of the Western Association of Fish and Wildlife Agencies (WAFWA) to assist state agencies in implementing creative and ``out of the box"  strategies, such as TYC, in the wild. Since the development of YY males for Brook Trout by the Idaho Department of Fish and Game in 2016/2017, these mutated fish have been released in several rivers and streams in Idaho, with promising results~\cite{schill2016production}, and the TYC strategy has become popular~\cite{dunker2022operational}. 

A key issue in implementing the TYC strategy is the production of supermales. Estrogen-induced feminization of the wild-type male fish is insufficient to obtain sex-reversed XY females, which would be required for the mass production of YY females. To overcome this issue, a harvesting-stocking strategy was proposed by Lyu \cite{lyu2020comparison}, where wild-type females are harvested while wild-type males are stocked. This strategy is called the FHMS strategy, and mimics the TYC dynamics. Other work has considered the FHMS strategy with weak Allee effects \cite{takyi2023gender}.

Discrete-time models are often underutilized in population dynamics studies, largely due to the challenges associated with identifying and analyzing equilibrium states. However, in scenarios such as the intermittent release of supermales into an ecosystem, discrete-time models provide a highly suitable framework. These models are particularly advantageous when the species’ life cycle involves distinct developmental stages, as they allow for a structured and stage-specific approach to modeling.

In a discrete-time framework, the species' life cycle is divided into several developmental stages. For each stage, a system of difference equations is formulated to capture the dynamics of transitions and interactions. These stage-specific subsystems are then integrated into a unified system of difference equations represented as
$$\mathbf x_{t+1} = F(\mathbf x_t),$$
where the vector $\mathbf x_t$ denotes the size of different sub-populations at time $t$ and $F(\mathbf x_{t})$ describes the transition dynamics across all stages. Typically, one time unit corresponds to the duration of a single life cycle, facilitating a natural temporal resolution for the population study.

One significant advantage of discrete-time models in this context is their capacity to simulate the release and impact of supermales with precision. Through simulations, researchers can control the frequency and quantity of supermales introduced per one release, enabling the exploration of various release strategies. This flexibility makes discrete-time models particularly effective for investigating optimal control strategies, allowing for the fine-tuning of interventions aimed at population management or ecological stabilization. By focusing on stage-specific dynamics and offering precise control over external interventions, discrete-time models serve as a powerful tool for understanding and managing population dynamics, especially in cases where life cycle stages and external controls play critical roles.

In the existing literature on TYC models, discrete-time models remain notably absent. This research aims to address this gap by developing and simulating an age-structured discrete-time model comprising two distinct life stages.
\section{Background}

The earliest TYC models that consist of a system of ordinary differential equations that accounts for the dynamics of three species:  adult females ($f$), adult males ($m$), and adult supermales ($s$) were proposed in \cite{parshad2011long, gutierrez2013global}. In these models, supermales are introduced into the system at a constant or density dependent rate $\mu(s)$. The classical model with constant introduction rate takes the following form,

\begin{eqnarray}
\label{ClassicTYCfeq} 
\begin{split}
\dot{f} &=&  \frac{1}{2}  \beta L fm - \delta f ,\\
\dot{m} &=&  \frac{1}{2}  \beta L fm + \beta L fs- \delta m,\\
\dot{s} &=&  \mu - \delta s,
\end{split}
\end{eqnarray}

Here, $\beta$ is the birth rate, $\delta$ is the death rate, $L = 1 - \dfrac{f + m + s}{K}$ implements logistic growth, where $K$ is the carrying capacity.

Spatially explicit as well as  stochastic versions of the TYC model have also been well studied, see e.g.,~\cite{wang2014analysis, parshad2011global, gutierrez2013global, wang2016stochastic}. The classical TYC model has since been shown to be ill-posed. For reference, key results established in \cite{takyi2022large} are paraphrased here:

\begin{lemm}

\label{lem:1a}
Consider the TYC system \eqref{ClassicTYCfeq}.
There exists positive initial data $(f_{0}, m_{0}, s_{0})$, and a time interval $[T_{1}, T_{2}] \in (0,\infty)$ such that for solutions emanating from this data, the male population $m(t)$ is negative on $[T_{1}, T_{2}]$.
\end{lemm}

\begin{thm}
\label{thm:2a}
Consider the TYC system \eqref{ClassicTYCfeq} with $\mu=0$.
Then there exists sufficiently large positive initial data $(f_{0}, m_{0}, s_{0})$ such that solutions emanating from this data will blow-up in finite time, that is
\begin{equation*}
\limsup_{t \rightarrow T^{*} < \infty} f(t) \rightarrow + \infty~~\text{or}~~\limsup_{t \rightarrow T^{**} < \infty} m(t) \rightarrow - \infty.
\end{equation*}

\end{thm}

\begin{thm}
\label{thm:2a1}
Consider the TYC system given by \eqref{ClassicTYCfeq} with $\mu>0$.
For any positive initial data $(f_{0}, m_{0}, s_{0})$, there exists a critical $\mu^{*}(f_{0}, m_{0}, s_{0})$, such that for any $\mu > \mu^{*}(f_{0}, m_{0}, s_{0})$, solutions emanating from this data, will blow-up in finite time, that is
\begin{equation*}
\limsup_{t \rightarrow T^{*} < \infty} f(t) \rightarrow + \infty~~\text{or}~~\limsup_{t \rightarrow T^{**} < \infty} m(t) \rightarrow - \infty.
\end{equation*}

\end{thm}

Due to these possible blow-up solutions, modifications to the classical TYC have also been considered \cite{takyi2023modified}. These include
\begin{itemize}
\item a modified logistic term for mating between supermales and females; 
\item female mating choice;
\item Intraspecific competition among males and supermales for females;
\item classical mating models that takes into account mating pairs.
\end{itemize}
The results for these modified models show that non-negative solutions, for any positive initial condition and parameters, exist globally in time. Also, they
 are effective in causing extinction of the invasive wild type.

\section{Materials and Methods}
We develop a two-stage age-structured discrete-time model to investigate the dynamics of a generic population that contains supermales. The two stages of the life cycle, namely egg and maturity are considered. Time for one life cycle is taken as one unit of time. The time intervals spent in egg and maturity stages are taken as $[t, t_1]$ and $[t_1, t+1]$ respectively. In other words, $t_1\in (0,1)$ describes the relative average time a fish spends in the egg stage over its life time. 

The fish population is stratified into the following five sub-populations based on age, gender and genotype:
\begin{itemize}[noitemsep]
    \item $x = $ adult females
\item $y = $ adult males
\item $z = $ supermales
\item $J_y = $ juvenile with male paternal DNA
\item $J_z = $ juveniles with supermale paternal DNA
\end{itemize}

To simplify the model, we assume that the beginning of the egg stage marks the starting point of the life cycle. Females are assumed to be mating with males and supermales randomly (i.e., we assume that the genotype has no impact on reproductive fitness). That implies the total number of eggs with male versus supermales paternal DNA is proportional to the male-supermale ratio. That is, the proportion of all eggs laid by one female after mating that have male paternal DNA ($f_y$) and supermale paternal DNA ($f_z$) is given by
\begin{align}
    f_y(y, z) &= \frac{y}{y + z}, \label{fecundity-M} \\ 
    f_z(y, z) &= \frac{z}{y + z}. \label{fecundity-S}
\end{align}


We introduce density dependence during the maturity stage. A biologically meaningful density dependent function $\rho$ is expected to satisfy the following conditions. 

\begin{description}
    \item[D1] $0\le \rho(w) \le 1$ for all $w \ge 0$.
    \item[D2] $\rho$ is a decreasing function.
    \item[D3] $\lim\limits_{w\to \infty}\rho(w)=0$ and $\rho(0) = 1$.
\end{description}

\textbf{D1} warrants that $\rho$ is the survival probability of the species. \textbf{D2} assures that when the population grows, the survival rate decreases. \textbf{D3} encodes natural extreme conditions of a survival probability.

In the following, we describe each stage of the life cycle in detail.
\begin{description}
    \item [Stage 1 (Egg):]
        During the period of $[t, t_1]$, adult females lay eggs, and after a certain amount of time the hatchlings become juveniles. During this period, $p_x$, $p_y$ and $p_z$ describe the survival probabilities of adult females, adult males and adult supermales, respectively. The fecundity $c$ describes the average number of offspring per female that survives through the egg stage and becomes a juvenile (at time $t_1$). The population sizes at time $t_1$ depend on the population sizes at time $t$ as follows:
            \begin{equation}\label{egg}
        \begin{split}
        	x(t_1) &= p_xx(t) \\
        	y(t_1) &= p_yy(t) \\
            z(t_1) &= p_zz(t) \\
            J_y(t_1) &= cf_y(y(t), z(t))x(t) \\
            J_z(t_1) &= cf_z(y(t), z(t))x(t)
        \end{split}
    \end{equation}
    See also Figure~\ref{fig:stages} (a).

\begin{figure*}[!h]
\centering
\includegraphics[width=0.7\textwidth]{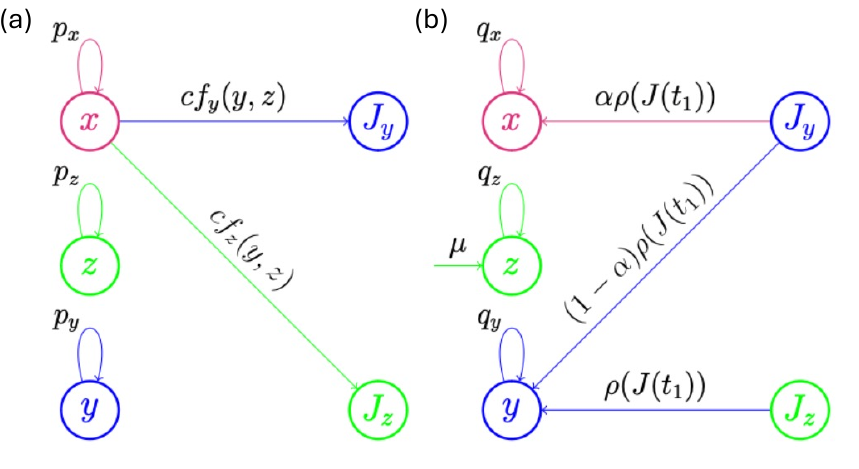}
\caption{Transitions between the different compartments at the two stages of the life cycle: (a) egg stage, (b) maturity stage. Here, $J(t_1) := J_y(t_1) + J_z(t_1)$ describes the argument of the density dependent juvenile survival function.}

\label{fig:stages}
\end{figure*}
    
    \item[Stage 2 (Maturity):]
    During this period of $[t_1, t+1]$, juveniles become adults. During this period, $q_x$, $q_y$ and $q_z$ are the survival probabilities of adult females, adult males and adult supermales, respectively. We assume that a fraction $\alpha$ of the surviving juveniles become adult females. Moreover, an amount of $\mu$ supermales are released into the system during this time. This can be described by the difference equations
            \begin{equation}\label{maturity}
        \begin{split}
        	x(t+1) &= q_xx(t_1) + \alpha \rho(J_y(t_1)+J_z(t_1))J_y(t_1),\\
        	y(t+1) &= q_yy(t_1) + \left(1-\alpha\right)\rho\left(J_y\left(t_1\right) 
            +J_z\left(t_1\right)\right)J_y\left(t_1\right) + \rho\left(J_y\left(t_1\right)
            +J_z\left(t_1\right)\right)J_z\left(t_1\right)\\
            z(t+1) &= q_zz(t_1) + \mu.
        \end{split}
    \end{equation}
    See also Figure \ref{fig:stages} (b).
\end{description}

Combining the two stages (i.e., eliminating the intermediate variable $t_1$ and the juvenile compartments from Eq. \eqref{egg} and Eq.\eqref{maturity}) yields a system of difference equations that accounts for the population dynamics of adult females, adult males and supermales over one entire life cycle. If we let $m_x := 1-p_xq_x$, $m_y := 1-p_yq_y$, $m_z := 1-p_zq_z$ describe the mortality rates, we have
    \begin{equation}\label{full-model}
        \begin{split}
        	x(t+1) &= (1-m_x)x(t) + \alpha c \rho(c x(t))f_y(y(t),z(t))x(t) \\
        	y(t+1) &= (1-m_y)y(t) + (1-\alpha) c \rho(c x(t)) f_y(y(t),z(t)) x(t) + c \rho(c x(t)) f_z(y(t), z(t)) x(t) \\
            z(t+1) & = (1-m_z) z(t) +\mu 
        \end{split}
    \end{equation}

\section{Equilibria}
The usual way to analyze a dynamical system is by finding its equilibria and investigating their stability. General closed-form expressions for the equilibria of the proposed model are impossible to find, especially because of the generic nature of the density dependent survival probability $\rho$. Nonetheless, the existence of equilibria of System \ref{full-model} and their stability can be investigated. 

We begin by considering the extinction equilibrium, where all ``regular" males and females have become extinct and the invasive species disappears once supermales are no longer introduced. 

\begin{thm}
    The extinction equilibrium of system (\ref{full-model}) always exists.
\end{thm}
\begin{proof}
    We find the equilibria by solving the following system of equations.
\begin{align*}
        x &= (1-m_x)x + \alpha c \rho(cx)f_y(y,z)x \\
        y &= (1-m_y)y + (1-\alpha) c \rho(cx) f_y(y,z)x + c \rho(cx) f_z(y,z)x \\ 
        z & = (1-m_z) z +\mu 
\end{align*}
If follows that
$$(x,y,z)=
\left(0,0,\dfrac{\mu}{m_z}\right)$$
is a solution, so an extinction equilibrium exists.
\end{proof}
For nonzero equilibria, we solve the following system.
\begin{align}
        1 &= 1 - m_x + \alpha c \rho(cx)f_y(y,z) \label{pos-equilibria-1} \\
        y &= (1-m_y)y + (1-\alpha) c \rho(cx) f_y(y,z)x + c \rho(cx) f_z(y,z)x \label{pos-equilibria-2}  \\ 
        z &= \dfrac{\mu}{m_z} \label{pos-equilibria-3}
\end{align}
From \eqref{pos-equilibria-2} we get,
\begin{equation}
   m_yy=((1-\alpha)f_y(y,z)+f_z(y,z))c\rho(cx)x.\label{intermediate 1}
\end{equation}
Substituting $f_z(y,z)=1-f_y(y,z)$ and simplifying we get,
\begin{equation}
    m_yy=c\rho(cx)x-\alpha c \rho(cx)x f_y(y,z) \nonumber
\end{equation}
and then eliminating $\alpha c \rho(cx)f_y(y,z)$ using \eqref{pos-equilibria-1}, we get
\begin{equation}
    m_yy=c\rho(cx)x - m_xx \label{intermediate 2}
\end{equation}
We substitute $f_y(y,z)=\dfrac{y}{y+z}$ and $z=\mu/m_x$ into \eqref{pos-equilibria-1} and solve for $y$ to get
\begin{equation}
    y=\dfrac{\mu m_x}{m_z(\alpha c \rho(cx) - m_x)}.\nonumber
\end{equation}
Substituting this expression for $y$ in \eqref{intermediate 2} and simplifying yields the following equation in $x$ only:
\begin{equation}
    \dfrac{\mu m_x m_y}{m_z} = (c\rho(cx)-m_x)(\alpha c\rho(cx)-m_x)x \nonumber
\end{equation}
Using the transformation $u=cx$, and rearranging the terms we get,
\begin{equation}
    \left(\dfrac{\mu m_xm_y}{\alpha c m_z}\right)\dfrac{1}{u} = \left(\rho(u)-\dfrac{m_x}{c}\right)\left(\rho(u)-\dfrac{m_x}{\alpha c}\right). \nonumber
\end{equation}
Finally, for the sake of simplicity, letting $\lambda = \dfrac{\mu m_x m_y}{\alpha c m_z}$ and $\beta = \dfrac{m_x}{c}$, we can rewrite this relationship in the form
\begin{equation}\label{key_equation}
    \dfrac{\lambda}{u} = \left(\rho(u)-\alpha \beta\right)\left(\rho(u)-\beta\right).
\end{equation}

For positive equilibria, we analyze this equation in the next section.

\subsection{Existence of a positive equilibrium}
We illustrate the existence of a positive equilibrium, using a graphical argument.
We show that the graphs of the two functions
$f(u)=\left(\rho(u)-\alpha \beta\right)\left(\rho(u)-\beta\right)$ and $g(u)={\lambda}/{u}$
intersect in the first quadrant.

\begin{figure}[hbt!]
\caption{Graph of $f(u)=\left(\rho(u)-\alpha \beta\right)\left(\rho(u)-\beta\right)$ for $ \alpha = 0.5, \beta =0.4, \mbox{ and }\rho(u) = \dfrac{1}{1 + u}.$}
\centering

\begin{tikzpicture}[x = 1pt, y = 1pt, scale = 1, every node/.style = {scale = 1}, scale = 1]

\begin{axis}[
    xmin = 0, xmax = 10,
    ymin = -0.05, ymax = 0.1,
    xlabel = $u$,
    ylabel = $f(u)$,
    grid = both,
    major grid style = {dotted,black},
    minor grid style = {dotted,black},
    scale = 0.8,
    minor tick num = 0,
    minor tick style = {draw=none},
    legend style = {draw = none, fill = none, at = {(0.1,1)}, anchor = north, legend cell align = left},
    width = 0.7\linewidth, no marks]
\addplot [line width = 0.5pt, solid, red, domain=0:10, samples=100] {(1/(1+x)-0.2)*(1/(1+x)-0.4)};

\end{axis}
\end{tikzpicture}
\label{fig: graph of f}
\end{figure}
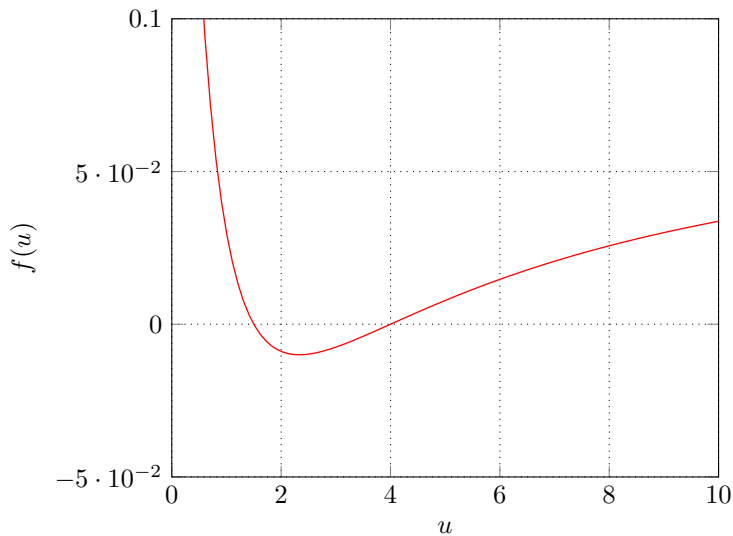

Recall that $\rho$ is a decreasing function such that $0\le \rho(u) \le 1$ for all $u\ge 0$ and $\lim\limits_{u\to \infty}\rho(u)=0$. Also note that $ 0 < \alpha \beta < \beta <1 $. We assume that fecundity $c > 1$, which means that on average breeding females lay enough eggs so that the average number of surviving juveniles per one female is greater than one. This is in fact the case for any species that has an increasing population. Figure \ref{fig: graph of f} shows a typical graph of $f$.

\begin{figure}[!h]
\includegraphics[width=\textwidth]{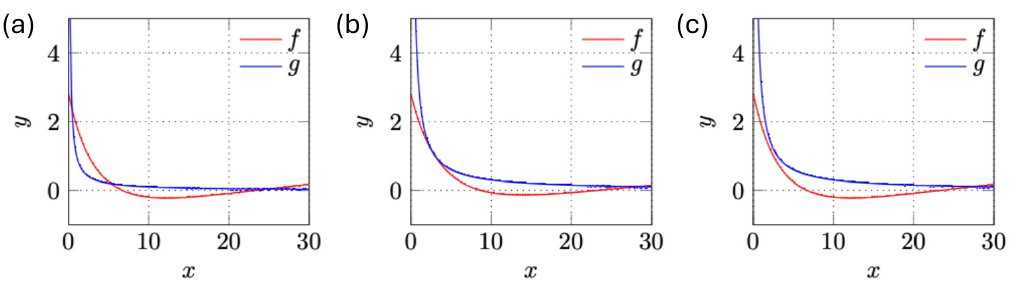}
\caption{Graphs of $f(x)=\left(\rho(x)-\alpha \beta\right)\left(\rho(x)-\beta\right)$ and $g(x)=\dfrac{\lambda}{x}$ with  (a) three, (b) two and (c) one  positive intersection point(s).}

\label{fig: f and g}
\end{figure}

Now, we graph the two functions $f(u)$ and $g(u)$ in the same plane. Note that,
\begin{align*}
    \lim\limits_{u\to \infty}f(u) &= \lim\limits_{u\to \infty} \left(\rho(u)-\alpha \beta\right)\left(\rho(u)-\beta\right)\\
     &= \alpha \beta^2 >0.
\end{align*}
Further, since $\rho(0) = 1$, we have
\begin{equation}
    f(0)=(1-\alpha \beta)(1-\beta)>0.
\end{equation}
Moreover,
\begin{equation*}
    \lim\limits_{u\to \infty}g(u) = \lim\limits_{u\to \infty}\dfrac{\lambda}{u}=0.
\end{equation*}
Since $\lim\limits_{u\to 0^+}g(u) = \infty$, the two graphs must intersect in at least one point in the first quadrant. Hence, as illustrated in Figure \ref{fig: f and g}, $f$ and $g$ can intersect in one, two or three points. Two intersections occur when the two graphs are exactly tangential to each other at one point. Thus, we have illustrated that a positive equilibrium exists, which we can also prove.

\begin{thm}
    A positive equilibrium of system (\ref{full-model}) always exists.
\end{thm}
\begin{proof}
    Let $h(u) = f(u) - g(u)$.
    
    Then, $\lim\limits_{u\to \infty}h(u) = \alpha\beta^2 >0$ and $\lim\limits_{u\to 0^+}g(u) < 0$.
    
    Hence, by continuity of $h$, the theorem follows.
\end{proof}

\subsection{Stability of the extinction equilibrium}
We prove the local stability of the extinction equilibrium.
\begin{thm}
    The extinction equilibrium of system (\ref{full-model}) is locally asymptotically stable.
\end{thm}
\begin{proof}
    The extinction equilibrium is $$(x_0,y_0,z_0)=
\left(0,0,\dfrac{\mu}{m_z}\right).$$
To find the Jacobian matrix $J$ for the System \eqref{full-model}, let
\begin{align*}
    F(x,y,z) &= (1-m_x)x+\alpha c \rho(cx)\dfrac{xy}{y+z}\\
    G(x,y,z) &= (1-m_y)y+(1-\alpha)c\rho(cx)\dfrac{xy}{y+z}+c\rho(cx)\dfrac{xz}{y+z}\\
    H(x,y,z) &= (1-m_z)z + \mu
\end{align*}
Then,
\begin{align*}
    \dfrac{\partial F}{\partial x} &= 1-m_x + \alpha c\left(c \rho^\prime(cx)\dfrac{xy}{y+z} + \rho(cx)\dfrac{y}{y+z}\right)\\
    \dfrac{\partial F}{\partial y} &= \alpha c \rho(cx)\dfrac{xz}{(y+z)^2}\\
    \dfrac{\partial F}{\partial z} &= -\alpha c \rho(cx)\dfrac{xy}{(y+z)^2}\\
    \dfrac{\partial G}{\partial x} &=(1 - \alpha)c\left(c\rho^\prime(cx)\dfrac{xy}{y+z} + \rho(cx)\dfrac{y}{y+z}\right) + c\left(c\rho^\prime(cx)\dfrac{xz}{y+z} + \rho(cx)\dfrac{z}{y+z}\right)\\
    \dfrac{\partial G}{\partial y} &= 1-m_y + (1 - \alpha) c\left(c\rho^\prime(cx)\dfrac{xy}{y+z}+\rho(cx)\dfrac{xz}{(y+z)^2}\right)\\
    \dfrac{\partial G}{\partial z} &= \alpha c\rho(cx)\dfrac{xy}{(y+z)^2}\\
    \dfrac{\partial H}{\partial x} &= 0\\
    \dfrac{\partial H}{\partial y} &= 0\\
    \dfrac{\partial H}{\partial z} &= 1-m_z
\end{align*}
Hence, the Jacobian matrix evaluated at $(x_0,y_0,z_0)=\left(0,0,\dfrac{\mu}{m_z}\right)$ is given by
\begin{equation*}
J(x_0,y_0,z_0)=
\begin{pmatrix}
    1-m_x & 0 & 0 \\
    c\rho(0) & 1-m_y & 0 \\
    0 & 0 & 1-m_z
\end{pmatrix}.
\end{equation*}
Being a lower triangular matrix, the eigenvalues are given by each of the main diagonal entries, namely $1-m_x$, $1-m_y$ and $1-m_z$, each of which is less than 1. Therefore, the local stability of the extinction equilibrium follows.
\end{proof}

If $\mu=0$, we have $z_0 = 0$ and supermales can be excluded. That implies we can consider the simplified model

\begin{align*}
    F(x,y) = F(x,y,z=0) &= (1-m_x)x+\alpha c \rho(cx)x\\
    G(x,y) = G(x,y,z=0) &= (1-m_y)y+(1-\alpha)c\rho(cx)x
\end{align*}
We find
\begin{align*}
    \dfrac{\partial F}{\partial x} &= 1-m_x + \alpha c\left(c \rho^\prime(cx)x + \rho(cx)\right)\\
    \dfrac{\partial F}{\partial y} &= 0\\
    \dfrac{\partial G}{\partial x} &= (1-\alpha) c\left(c \rho^\prime(cx)x + \rho(cx)\right)\\
    \dfrac{\partial G}{\partial y} &= 1-m_y
\end{align*}
Hence, the Jacobian matrix evaluated at $(x_0,y_0)=\left(0,0\right)$ is given by
\begin{equation*}
J(x_0,y_0)=
\begin{pmatrix}
    1-m_x + \alpha c\rho(0) & 0 \\
    (1-\alpha) c\rho(0) & 1-m_x
\end{pmatrix}.
\end{equation*}
Being a lower triangular matrix, the eigenvalues are given by each of the main diagonal entries, namely $1-m_x + \alpha c\rho(0)$ and $1-m_y$. We have $1-m_y<1$. However, the zero steady state is locally asymptotically stable if 
\begin{equation}
    \label{stable_zero_steady_state}
    m_x > \alpha c\rho(0). 
\end{equation}

This inequality makes sense intuitively. For a tiny population of size $\epsilon > 0$, extinction is unavoidable if the rate at which females die ($m_x$) is larger than the rate at which females produce new female offspring that survives to adulthood ($\alpha c \rho(\epsilon) \approx \alpha c \rho(0)$).

\subsection{Numerical Simulations}
The complexity of the full model prevents a comprehensive theoretical analysis. We therefore resort to simulations, where we assume a simple function for density dependence. We use a hyperbolic decay function defined by
\begin{equation}\label{hyperbolic}
    \rho(x)=\frac{K}{K+x},
\end{equation}
where we set $K = 100$. Notice that this choice of $\rho$ satisfies the properties D1-D3 given in the model formulation.

We further choose the following default values: for the survival probabilities, we set $p_i = 0.6, q_i = 0.7$ so that the mortality rate is $m_i = 0.58$ for all three types of adults $i \in \{x,y,z\}$. By default, we set $\alpha=0.5$ (i.e., we assume a 1:1 sex ratio), we choose a value of $c=4$ (low) or $c=20$ (high) for the fecundity, and we assume $\mu=100$ supermales are released each life cycle. 

\begin{figure}
    \centering
    \includegraphics[width=\linewidth]{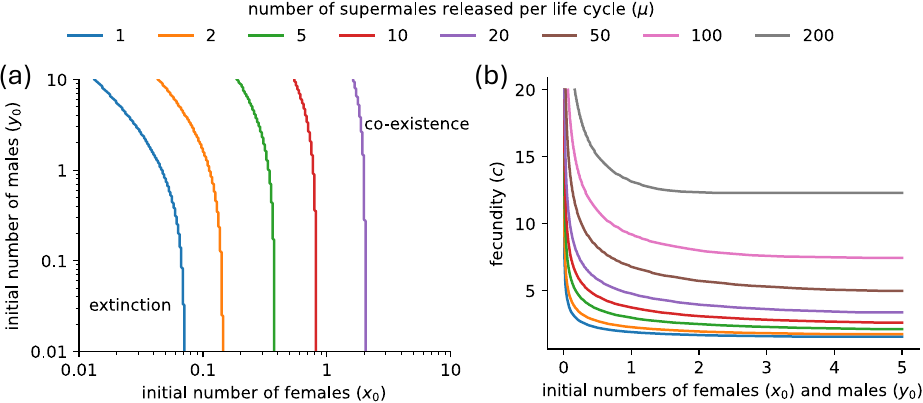}
    \caption{Bistable model behavior. For different numbers of supermales ($z$) released per life cycle ($\mu$; colors; $z_0=0$), the lines in (a) show the separatrix between the region of initial numbers of females ($x_0$; x-axis) and males ($y_0$; y-axis) that result in extinction (towards the origin) and co-existence (away from the origin). The fecundity is set to $c=4$. In (b), equal initial numbers of males and females are assumed (i.e., $x_0 = y_0$; x-axis) and the fecundity is varied (y-axis). The lines describe again where the switch from co-existence (away from the origin) to extinction (towards the origin) occurs.}
    \label{fig:separatrix}
\end{figure}

An analysis of the model dynamics for different initial conditions reveals that the qualitative long-term behavior depends heavily on the initial number of females~(Fig.~\ref{fig:separatrix}a). If this number is sufficiently high, extinction does not occur. The required number of females depends strongly on the number of supermales that are released per life cycle and, to a much lesser extent, on the initial number of males that can compete with the supermales as mating partners. If the number of added supermales is too high, extinction is unavoidable if the fecundity is rather low (e.g., $c=4$ as in Fig.~\ref{fig:separatrix}a). Fecundity, the number of periodically released supermales, and the initial numbers of males and females collectively determine the long-term outcome~(Fig.~\ref{fig:separatrix}b). As expected, the higher the fecundity, the more supermales must be added each life cycle to ensure extinction of the invasive species. Here, the initial population size, unless tiny, proved less important. 

To increase realism and the applicability of the model results, we assume from here on that the introduction of supermales starts at a random point in time and that the system, in the absence of supermales, rests at a positive equilibrium. For the function $\rho$ given in \eqref{hyperbolic}, it is straight-forward to see that System~\ref{full-model}, in the absence of supermales, only possesses a single positive steady state, given by
\begin{equation}
    x^* = \frac{K(\alpha c - m_x)}{cm_x}, \quad y^* = \frac{(1-\alpha) c K x^*}{m_y(K+cx^*)}, \quad z^* = 0.
\end{equation}
Note that $x^* > 0$ whenever $\alpha c > m_x$, which, since $\rho(0) = 1$, coincides exactly with the range of parameters where the extinction steady state is unstable~(Eq.~\ref{stable_zero_steady_state}). 

\begin{figure}
    \centering
    \includegraphics[width=\linewidth]{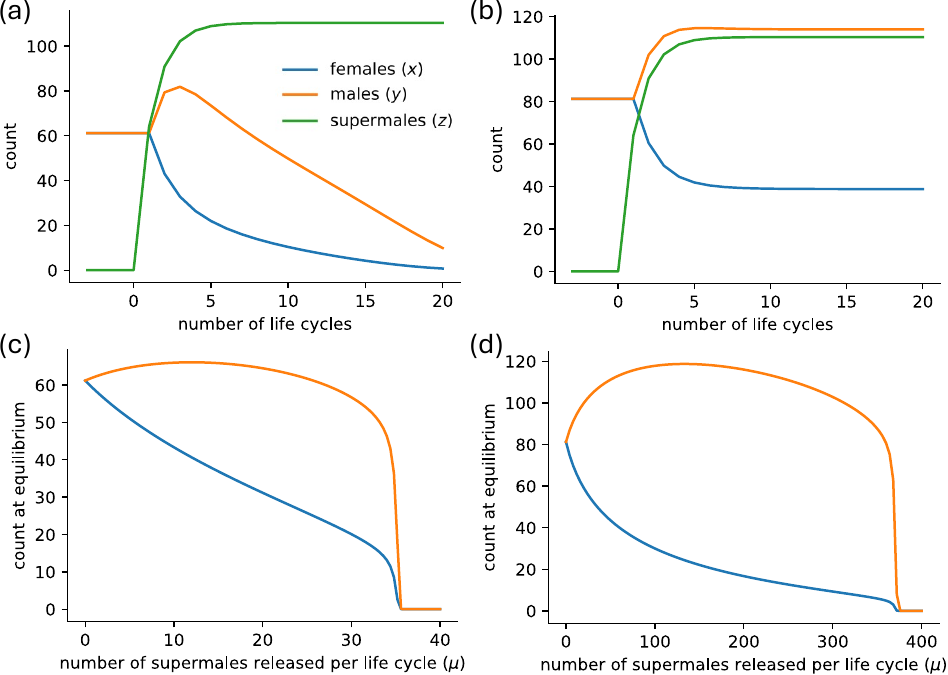}
    \caption{(a-b) Time series of population numbers after the continuous addition of $\mu=100$ supermales per life cycle. (c-d) Equilibrium population numbers for different choices of $\mu$. 
    The fecundity (i.e., the number of eggs per female that survive into juvenile stage) is set to $c=4$ in (a,c) and $c=20$ in (b,d). All other parameters are at their default values.}
    \label{fig:timeseries}
\end{figure}

When supermales are introduced, their population numbers quickly rise and their sub-population dynamics and equilibrium $z^*$ only depend on the introduction rate and the mortality rate: $z^* = \mu/m_x$. The presence of supermales always leads to an immediate reduction in the number of females (Fig.~\ref{fig:timeseries}a,b). The effect on males is more complicated: If the fecundity is low (e.g., $c=4$ as in Fig.~\ref{fig:timeseries}a), females and thus males both eventually vanish, with the male population exhibiting a short-lived increase in population numbers right after the introduction of supermales. This is due to the fact that offspring with supermale paternal DNA is always male. If the fecundity is higher (e.g., $c=20$ as in Fig.~\ref{fig:timeseries}b), a sufficiently high number of females continues to mate with regular males so that the female population (and thus the male population) settles at a positive steady state. 

The constant introduction of $\mu=100$ supermales only causes extinction if the fecundity is sufficiently low. Variation of the parameter $\mu$ reveals that extinction occurs whenever $\mu\geq 33.6$ if $c=4$ (Fig.~\ref{fig:timeseries}c) and whenever $\mu\geq 360$ if $c=20$ (Fig.~\ref{fig:timeseries}d). If fewer supermales are introduced, the ``regular" population will persist, albeit at lower total population numbers. Because female-supermale mating produces only males, equilibrium numbers for males must always be higher than for females (as long as $m_x\geq m_y$ and $\alpha\leq 0.5$). Interestingly, for a large range of lower $\mu$-values, the male sub-population numbers rise to a level higher than in the absence of supermales.

\begin{figure}
    \centering
    \includegraphics[width=\linewidth]{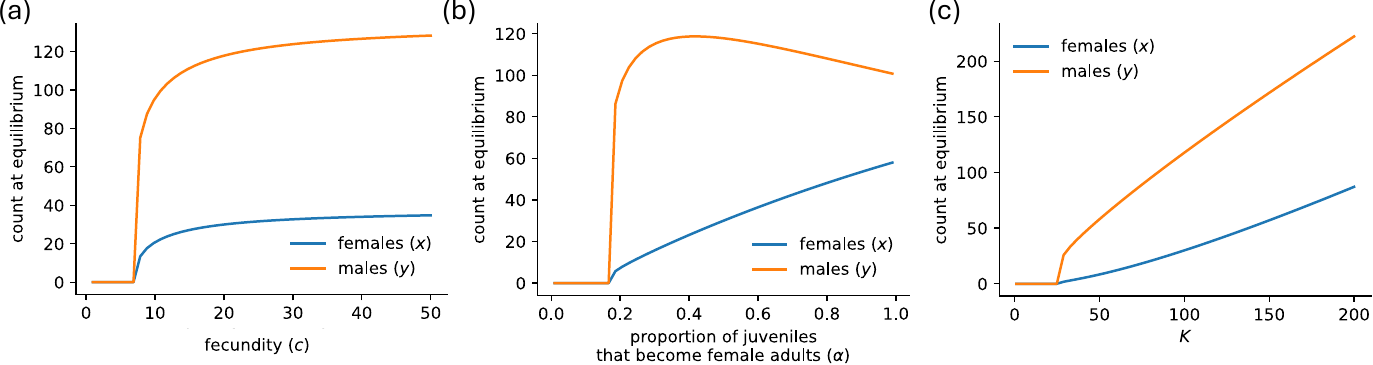}
    \caption{Equilibrium population numbers for different choices of (a) the fecundity $c$, (b) the proportion of juveniles that become female adults $\alpha$, and (c) the parameter $K$ of the function $\rho$ in (Eq.~\ref{hyperbolic}).  All other parameters are at their default values. Specifically, the fecundity in (b) and (c) is $c=20$.}
    \label{fig:other_bifurcations}
\end{figure}

\begin{figure}
    \centering
\includegraphics[width=\linewidth]{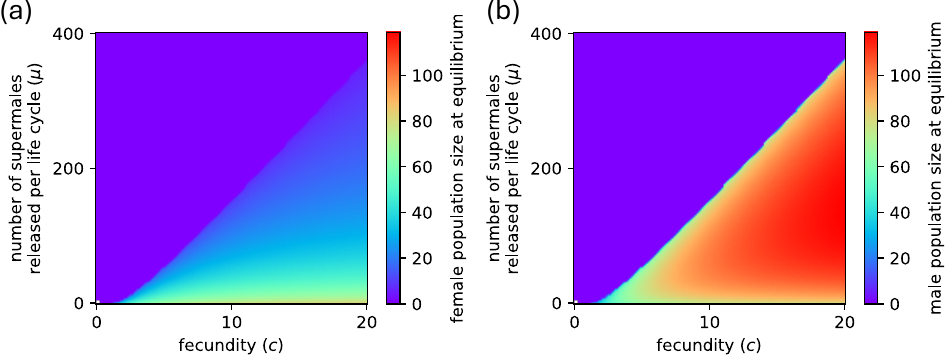}
    \caption{(a) Female and (b) male equilibrium population numbers for different combinations of fecundity (c) and released supermales per life cycle ($\mu$). All other parameters are at their default values.}
    \label{fig:bif_2D_heatmap_mu_c}
\end{figure}

As the fecundity increases beyond the point where the positive equilibrium becomes stable, the equilibrium population numbers of males increase much faster and to much higher levels than those of females~(Fig.~\ref{fig:other_bifurcations}a). This is even more prevalent when varying both the fecundity ($c$) and the number of released supermales ($\mu$; Fig.~\ref{fig:bif_2D_heatmap_mu_c}). These bifurcation plots also reveal an approximately linear bifurcation line in $c$-$\mu$ space. They further reveal that the finding of male equilibrium population numbers that are non-monotonic in $\mu$ -- already observed for $c=4$ and $c=20$ in Fig.~\ref{fig:timeseries}c,d --  persists for a large range of fecundity values. On the contrary, female equilibrium population numbers linearly decrease as the number of released supermales increases, for all fecundity values.

The rapid increase in male equilibrium population numbers can also be seen when the proportion of juveniles that become female adults ($\alpha$), which we kept at $50\%$ thus far, is varied. As $\alpha$ is increased, female equilibrium population numbers rise slowly, while male numbers rise very quickly before they decrease as $alpha$ is further increased, favoring the emergence of females. Variation of the parameter $K$ of the survival function $\rho$ given in \eqref{hyperbolic} for juvenile yields similar, albeit fully monotonic results: As $K$ is increased, both the male and female equilibrium numbers steadily increase, with male numbers rising faster.

\section{Discussion}
This study focuses on understanding how the introduction of supermales in the context of the Trojan Y Chromosome (TYC) strategy can affect the population dynamics of invasive species. We developed an age-structured discrete mathematical model to study this effect, considering two main stages of the life cycle: the egg stage and the maturity stage. The model accounts for the impact of releasing genetically modified males (supermales), which harbor only Y chromosomes so that their offspring can only be male. This explains why the introduction of supermales shifts the population toward more males, which can eventually lead to the extinction of the invasive species.

The discrete-time approach is suitable when supermales are released at specific times. The mathematical analysis shows that extinction of the invasive species is possible under certain conditions, especially when a large enough number of supermales can be introduced. The minimal number of supermales that must be introduced each life cycle to ensure extinction is strongly influenced by the fecundity of the adult females.

Stability analysis shows that the general model possesses between one and three positive equilibria. Numerical simulations indicate that any of these positive equilibria are only locally asymptotically stable: If initial population numbers are too small, extinction is unavoidable.

In this first discrete-time model describing the TYC strategy, we assumed that the number of released supermales remains constant over time. In practice, this certainly does not need to be the case. This naturally gives rise to interesting optimal control problems such as: Minimize the cost (and potentially the time to extinction) under the constraints that extinction must be achieved and that limited, possibly time-varying maximal quantities of supermales are available. Meaningful answers to such questions require the availability of accurate parameters related to costs and constraints. 

Overall, this new model provides useful insights into how the TYC strategy can be studied and implemented more effectively. It offers a practical framework for planning and decision-making in the control of invasive species and helps fill a gap in current research.
\bibliographystyle{apalike}

\end{document}